\documentclass[aps,showpacs]{revtex4}

\def\RC{{\rm RC}}
\def\gAq{g_A(q_0^2)}
\def\gpq{g_P(q_0^2)}
\def\nn{\nonumber \\}
\def\He3{ { }^3{\rm He}}
\def\captHe{\mu^-\, \He3 \to \nu_\mu\, ^3{\rm H}}
\def\muHe3{\mu^-\, \He3}
\def\tbound{\tau_\mu^{\rm bound}}
\def\tfree{\tau_\mu^{\rm free}}
\def\capt{\mu^- N\to \nu_\mu N'}
\def\captH{\mu^- p\to \nu_\mu n}
\def\gpex{g_P^{\rm exp}}
\def\gpth{g_P^{\rm theory}}

\def\order#1{{\cal O}\!\left(#1\right)}
\newcommand{\ba}{\begin{eqnarray}}
\newcommand{\ea}{\end{eqnarray}}
\newcommand{\beq}{\begin{equation}}
\newcommand{\eeq}{\end{equation}}

\begin{document}

\title{Electroweak Radiative Corrections to Muon Capture}

\author{Andrzej Czarnecki}
\affiliation{Department of Physics, University of Alberta\\
Edmonton, AB\ \  T6G 2J1, Canada}

\author{William J. Marciano}
\affiliation{Brookhaven National Laboratory,
Upton, NY 11973, USA }

\author{Alberto Sirlin}
\affiliation{Department of Physics, New York University,\\
    4 Washington Place, New York, NY 10003, USA}

\begin{abstract}
Electroweak radiative corrections to muon capture on nuclei are
computed and found to be sizable.  They enhance the capture rates for
hydrogen and helium by 2.8\% and 3.0\% respectively.  As a result, the
value of the induced pseudoscalar coupling, $\gpex$, extracted from a
recent hydrogen $1S$ singlet capture experiment is increased by about
21\%
to $\gpex = 7.3\pm 1.2$ and brought into good agreement with the
prediction of chiral perturbation theory, $\gpth=8.2\pm 0.2$.
Implications for helium capture rate predictions are also discussed.

\end{abstract}


\maketitle
The study of muon capture by nuclei, $\mu^- N\to \nu_\mu N$, has
played an important role in the development of weak interaction
physics \cite{Primakoff:1959fs,Mukhopadhyay:1976hu}.  Used primarily
in the past to explore nuclear structure and its effects on weak
interactions, muon capture can now be employed to test Quantum
Chromodynamics (QCD) and its basic chiral symmetries
\cite{Gorringe:2002xx}.  In addition, it can provide a possible window
or constraint on new high mass scale physics \cite{Govaerts:2000ps},
beyond Standard Model expectations, such as additional gauge bosons,
charged Higgs scalars, leptoquarks etc.  Of course, to be competitive
with other precision low energy experimental tests of the Standard
Model, both theory and experiment for muon capture must be known to a
fraction of a percent.

Here, we would like to advance the theory of muon capture to that high
level of precision by including Standard Model electroweak radiative
corrections and estimating their degree of reliability.  From our
previous work
\cite{Sirlin:1974ni,Sirlin:1978sv,Marciano:1986pd,Czarnecki:2004cw,Marciano:2005ec}
on neutron (and nuclear) $\beta$-decay, one can anticipate that such
quantum loop effects are relatively large, $\sim 2-3\%$, and therefore
important for any precision confrontation between muon capture theory
and experiment.  As we shall show, that indeed is the case.

We begin by recalling the basics of muon capture.  Negative muons,
$\mu^-$, are stopped in matter.  They bind electromagnetically with
nuclei and quickly cascade down to the lowest energy atomic orbitals.
There, primarily from $1S$ states, the muon's final fate is to undergo
either ordinary muon decay, $\mu^- \to e^- \bar \nu_e \nu_\mu$, or
weak capture $\mu^- N\to \nu_\mu N'$ on the nucleus.  

Ordinary decay in orbit occurs essentially at the same rate as in
vacuum (modulo bound state time dilation and other small effects
\cite{ueberall60,Czarnecki:1999yj}).
The already well known "free" muon lifetime has been recently
remeasured \cite{MuLan2007}, 
thereby leading to the improved world average
\begin{equation}
              \tau_\mu =2.197019(21) \times 10^{-6} \,\sec
\label{eq:1}
\end{equation}
Further improvement by an additional factor of ten is expected.

The competing weak capture reaction, $\mu^- N\to \nu_\mu N'$, proceeds
via $W$ boson exchange with the nucleus.  Due to an
overlap flux factor from its atomic wavefunction at the origin squared
and a factor of $Z$ (nuclear charge) corresponding to the number of
nuclear protons that can induce capture, the overall capture rate
scales very roughly as $Z^4$.  In hydrogen ($Z=1$), the capture rate
is predicted to be very small.  From the $1S$ singlet (spin 0)
$\mu^-p$ state, it is only about $0.16\%$ of the ordinary decay rate
and for the triplet (spin 1) bound state configuration, it is a tiny
$0.0025\%$.  Those small rates make experimental  hydrogen capture
studies difficult, which is unfortunate, since hydrogen theory is very clean.
Decay and capture rates become comparable for $Z\simeq 10$, while at
much higher $Z$, capture dominates.

An interesting technique used to obtain muon capture rates involves
comparing free and bound $\mu^-$ lifetimes,
\begin{equation}
\Gamma(\capt) = {1\over \tbound} - {1\over \tfree},
\label{eq:2}
\end{equation}
(after making small bound state lifetime corrections).  Using an
ingenious application of that lifetime technique, the MuCap collaboration
\cite{Andreev:2007wg} at PSI recently reported a precise measurement of the 1S
singlet capture rate in hydrogen,
\begin{equation}
\Gamma(\captH)_{1S}^{\rm singlet} = 725.0\pm 13.7(\rm stat) \pm 10.7
(\rm sys) /{\rm sec}.
\label{eq:3}
\end{equation}
That already impressive $\pm 2.4\%$ level  of accuracy is expected to 
further improve to better than $\pm 1\%$ as additional data is
analyzed.

In the case of helium, the capture rate for $\captHe$
has been even better measured by directly detecting the charged
final state $ ^3{\rm H}$.  For the statistical combination of singlet (spin
0) and triplet (spin 1) $1S$ $\muHe3$ bound states,
\begin{equation} 
\Gamma(\captHe)_{\rm stat}=
{1\over 4}\Gamma(\captHe)_{\rm singlet}
+{3\over 4}\Gamma(\captHe)_{\rm triplet},
\label{eq:4}
\end{equation}
a long standing result \cite{Ackerbauer:1997rs}
\begin{equation} 
\Gamma(\captHe)_{\rm stat}^{\rm exp} = 1496(4)/\sec,
\label{eq:5}
\end{equation}
represents a remarkable $\pm 0.3\%$ determination.  

The Standard Model theoretical prediction for the basic $\captH$
capture rate depends on four relativistic form factors that result
from nucleon matrix elements of the $V-A$ weak quark charged current,
\begin{eqnarray}
\langle n |\bar d \gamma_\alpha (1-\gamma_5) u|p\rangle
&=&
\bar u_n(p_2) \left[
F_1(q^2)\gamma_\alpha
+{i\over 2m_N} F_M(q^2) \sigma_{\alpha\beta}q^\beta
-g_A(q^2)\gamma_\alpha\gamma_5
-{1\over m_\mu} g_P(q^2)q_\alpha\gamma_5 
\right]
u_p(p_1),
\nn
q&\equiv & p_2-p_1, \quad m_N \equiv {m_p+m_n\over 2}.
\label{eq:6}
\end{eqnarray}
Two other form factors, scalar and pseudotensor, are in general
possible, but are negligibly small in the Standard Model (arising from
isospin violation).  They should, however, be included in general
searches for ``New Physics'' effects \cite{Govaerts:2000ps,Santisteban:1976xt}.
In terms of the above form factors, the capture rate is given by  (modulo
radiative corrections, discussed later)
\begin{eqnarray}
\left. \Gamma(\captH)\right|_{\rm singlet} &=& |\psi(0)|^2 
{G_\mu^2 |V_{\rm ud}|^2 
\over 
2\pi}
{ E_\nu^2 \over M^2} (M-m_n)^2
\nn
&& 
\cdot \left\{
{2M-m_n\over M-m_n} F_1
+{2M+m_n\over M-m_n} g_A
-{g_P\over 2}
+ \left(2 M+2 m_n-3m_\mu\right){F_M\over 4m_N}
\right\}^2,
\nn
\left. \Gamma(\captH)\right|_{\rm triplet} &=& |\psi(0)|^2 
{G_\mu^2 |V_{\rm ud}|^2 
\over 
24\pi}
{ E_\nu^2\over M^2}(M-m_n)^2
\nn
&& 
\cdot \left\{
 \left[
 g_P 
-{2 m_n \over M-m_n}  \left(F_1-g_A\right) 
+ \left(2 M+2 m_n-m_\mu\right)  {F_M\over 2m_N}
\right]^2
\right.
\nn
&&
\left.
\quad 
+2 \left[
 g_P 
+ {2 M\over M-m_n}  \left(F_1-g_A\right) 
-m_\mu  {F_M\over 2m_N}
\right]^2  
\right\} .
\label{eq:1x}
\end{eqnarray}
$M$ denotes the mass of the $\mu^- p$ atom. We neglect the
binding effect and use $M\equiv m_p+m_\mu$.
The $\mu^- p$ hydrogenic  wave function at the origin is 
\begin{eqnarray}
|\psi(0)|^2 &=& {\mu^3 \alpha^3 \over \pi}\left(1-4 \alpha \mu r_p
\right) \simeq {\mu^3 \alpha^3 \over \pi}\left(1-0.005 \right) ,
\nn
\mu &\equiv & {m_p m_\mu \over m_p + m_\mu } \quad \mbox{(reduced
 mass)},
\label{eq:2x}
\end{eqnarray}
where we have accounted for the proton charge distribution with the
radius $r_p = {0.862 \over \sqrt{6} }$ fm (see
\cite{protonRadius862,Pachucki1999Proton} for a more detailed
discussion).

Three of the four form factors in Eq.~(\ref{eq:6}) are very well
determined at $q^2=0$ from CVC and neutron $\beta$ decay
\cite{Czarnecki:2004cw,PDG2006}, 
\begin{eqnarray}
F_1(0) &=& 1,\nn
F_M(0) &=& 3.706,\nn
g_A(0) &=& 1.2695(29).
\label{eq:7}
\end{eqnarray}
Extrapolating to $q_0^2=-0.88m_\mu^2$, as appropriate for $\mu^-$
capture on hydrogen, one finds
\begin{eqnarray}
F_1(q_0^2) &=& 0.976(1),\nn
F_M(q_0^2) &=& 3.583(3),\nn
g_A(q_0^2) &=& 1.247(4),
\label{eq:8}
\end{eqnarray}
where the errors include estimated $q^2$ evolution uncertainties.  

In the case of the induced pseudoscalar coupling, $\gpq$, PCAC
(partially conserved axial current) and chiral perturbation theory
predict \cite{Kaiser:2003dr,Bernard:1994wn,Adler:1966gc,Wolfenstein1970}
\begin{equation}
\gpq = {2m_\mu g_{\pi pn}(q_0^2) F_\pi \over m_\pi^2 -q_0^2 }
-{1\over 3} g_A(0) m_\mu m_N r_A^2,
\label{eq:9}
\end{equation}
which for $g_{\pi pn}=13.05(20)$, $F_\pi=92.4(4)$ MeV, and $r_A^2 =
0.43(3)\,{\rm fm}^2$ implies
\begin{equation}
\gpq =8.2\pm 0.2.
\label{eq:10}
\end{equation}
That prediction is expected to be very reliable, depending only on the
chiral properties of QCD and principles of PCAC. Nevertheless, it
would be very useful to have a first-principles lattice QCD calculation
of $\gpq$ (as well as $\gAq$).  Of course, it is also very important
to verify the prediction in Eq.~(\ref{eq:10}) experimentally.

Employing the above form factors at $q_0^2$ and allowing for the
variation $\gpq=8.2 +\delta g_P$, one obtains from Eq.~(\ref{eq:1x}) 
the singlet $1S$ capture rate on hydrogen,
\begin{equation}
\Gamma(\captH)_{1S}^{\rm singlet} =
692.3(3.4)\left(1+\RC({\rm H})\right)(1-0.0108\delta g_P)^2/\sec .
\label{eq:11}
\end{equation}
$G_\mu = 1.166371(6)\times 10^{-5}\, {\rm GeV}^{-2}$ (the Fermi
constant obtained from the free muon lifetime \cite{MuLan2007}), 
$V_{ud}=0.9738$ and a
0.5\% reduction from the finite proton size have been incorporated
into Eq.~(\ref{eq:11}).  The $1+\RC({\rm H})$
factor represents the effect of electroweak radiative corrections,
which up until this work have not been seriously considered in
discussions of muon capture 
\cite{Bernard:2000et,Ando:2000zw}.  If we
set $\RC({\rm H})=0$ and compare Eq.~(\ref{eq:11}) with
Eq.~(\ref{eq:3}), we find $\gpq =6.0 \pm 1.2$ which is about $2\sigma$
below the prediction in Eq.~(\ref{eq:10}); however, that result is not
very meaningful since we expect the radiative corrections to be 
sizable.

In the case of helium, the tree level theoretical prediction for muon
capture is not as pristine. When compared with the same input
parameters, two distinct approaches
give somewhat different results.  The first is based on an
elementary particle prescription which treats $\He3$ and $ ^3{\rm H}$
as initial and final particle states \cite{FujiiYama,Gorringe:2002xx}.  
It then
employs form factors analogous to those in Eq.~(\ref{eq:6}) (but
defined
with
an additional minus sign for all but $F_1$) at $q^2 = -0.954 m_\mu^2$
appropriate for $\mu^-$ capture on $\He3 \to ^3{\rm H}$.  Using CVC
for the vector form factors and PCAC to relate axial-vector and
pseudoscalar form factors, the analysis 
leads to what has been viewed as a rather
reliable $\He3$ capture rate prediction.  It depends primarily on the
input
\begin{equation}
g_A(q^2 = -0.954 m_\mu^2)_{\He3\to  ^3{\rm H}} = 1.052\pm 0.005,
\label{eq:12}
\end{equation}
obtained by evolving $g_A(0)_{\He3\to ^3{\rm H}} = 1.212$, obtained
from tritium $\beta$ decay \cite{Budick:1991zb,Simpson87}, to
$q^2=-0.954 m_\mu^2$.

The second method for calculating the capture rate for $\captHe$ uses
an impulse approximation to combine the basic $\captH$ captures
within $\He3$ \cite{Congleton:1993,Gorringe:2002xx}.  
It has been argued that when
supplemented by meson exchange current corrections
\cite{Congleton:1995db}, this method agrees with the above (elementary
particle) approach.  However, a close scrutiny of the most detailed
impulse approximation study \cite{Marcucci:2001qs} reveals some
difference in their predictions.

Normalizing to $V_{\rm ud}=0.9738$, the elementary particle model
approach predicts   \cite{Congleton:1993}
\begin{equation}
\Gamma\left( \captHe \right)^{\rm EPM}_{\rm stat} = 1492 (21)\cdot
(1+\RC({\rm He})) /\sec,
\label{eq:26}
\end{equation}
while the impulse approximation study by Marcucci {\em et al.}
\cite{Marcucci:2001qs} updated to a central value of $g_A=1.2695(29)$
and $g_P=8.2 + \delta g_P$
gives
\begin{equation}
\Gamma\left( \captHe \right)^{\rm IA}_{\rm stat} = 1462(8)(7)_{g_A}
\cdot
(1+\RC({\rm He}))
(1-0.013 \delta g_P)
 /\sec,
\label{eq:27}
\end{equation}
Again, we have allowed for inclusion of electroweak radiative
corrections, $\RC({\rm He})$, appropriate for capture.  For clarity,
we note that the values reported by Marcucci {\em et al.}
\cite{Marcucci:2001qs} are larger than in Eq.~(\ref{eq:27}) because
these authors identified $G_V^2$ with a parameter $G_V'^2\equiv
1.024 |V_{ud}|^2 G_\mu^2$, extracted from superallowed beta decays, in
which inner radiative corrections of 2.4\% were already included. In
Eq.~(\ref{eq:27}), we have factored out this 2.4\% effect and included
it in the overall RC(He) to be discussed below.

The prediction in Eq.~(\ref{eq:26}) is in very good agreement with
Eq.~(\ref{eq:5}), $\Gamma(\captHe)_{\rm stat}^{\rm exp} =
1496(4)/\sec$, if we naively set $\RC({\rm He}) =0$.  That agreement
has been viewed as a success of theory and used to constrain
\cite{Govaerts:2000ps} ``New Physics'' appendages to the Standard
Model.  On the other hand, Eq.~(\ref{eq:27}) only agrees with
experiment if one includes the +2.4\% radiative correction
contained in their $G_V^2$ value.

Now, we consider the electroweak radiative corrections (RC).  They
naturally divide into two contributions.  The first set is essentially
common to all semileptonic weak charged current amplitudes normalized
in terms of $G_\mu$, the  Fermi constant obtained from the free muon
lifetime.  The second type of correction stems from QED corrections to
the muonic atom wavefunction.  As pointed out by Goldman
\cite{Goldman:1973qu}, those latter effects are dominated by vacuum
polarization corrections to the Coulombic bound state interaction.

Making the above division,
\begin{equation}
\RC(N)=\RC(N)_1+\RC(N)_2,
\label{eq:14}
\end{equation}
we find from the detailed studies of neutron decay
\cite{Czarnecki:2004cw,Marciano:2005ec} 
(neglecting terms of
relative order $\alpha m_\mu/m_N$) that the $\order{\alpha}$ electroweak
radiative corrections  to the muon capture rate on hydrogen are given by 
\begin{equation}
\RC({\rm H})_1 = {\alpha \over 2\pi} \left[
4\ln {m_Z\over m_p} -0.595 +2C +g(m_\mu, \beta_\mu=0)\right],
\label{eq:15}
\end{equation}
where $m_Z=91.1875$ GeV, $m_p =0.938$ GeV, 
\begin{equation}
C=0.829,
\label{eq:16}
\end{equation}
and the quantity $g(m_\mu, \beta_\mu=0)$ can be obtained from
Eq.~(20b) in Ref.~\cite{Sirlin67} by replacing $m_e\to m_\mu$,
ignoring bremsstrahlung and taking the non-relativistic (zero muon
velocity) $\beta_\mu=0$ limit.  In that way one finds
\begin{equation}
g(m_\mu, \beta_\mu=0)=3\ln{m_p\over m_\mu} -{27\over 4} =-0.199.
\label{eq:17}
\end{equation}
In total, Eq.~(\ref{eq:15}) gives 0.0223.  Summing up higher order
leading logs along the lines of
ref.~\cite{Marciano:1986pd,Czarnecki:2004cw} 
enhances
that correction somewhat to 
\begin{equation}
\RC({\rm H})_1 =0.024(4),
\label{eq:18}
\end{equation}
where we have included a fairly generous estimate of the uncertainty.
It corresponds to roughly a $\pm 100\%$ variation in $C$ and
conservatively allows for $\order{\alpha m_\mu/m_p}$ corrections that
we have not computed.

We note that the first two bracketed terms in Eq.~(\ref{eq:15}) (which
include QCD perturbative effects) are of short-distance origin and
therefore apply to all muon capture rates.  Similarly the $g$ function
is essentially unchanged as long as the muon is non-relativistic and
$\order{\alpha m_\mu/m_p}$ contributions are ignored.  On the other
hand, the quantity $C$ in Eq.~(\ref{eq:16}) is specific to hydrogen
and will be modified by nucleon interactions in multi-nucleon systems.
Rather than try to account for that modification, we assume that our
rather conservative error covers those variations and continues to hold,
\begin{equation}
\RC({\rm He})_1 = 0.024(4),
\label{eq:19}
\end{equation}
If needed, the correction in Eq.~(\ref{eq:19}) can be used as a good
approximation for any muon capture rate.  We note that our +2.4\%
correction happens to coincide numerically with the inner radiative
corrections included in the $G_V^2$ value employed in
\cite{Marcucci:2001qs}.

At this point we note that the factorization of the radiative
corrections comes about because in the formulation of
Ref.~\cite{Czarnecki:2004cw}, which we follow, the axial couplings in
neutron decay have by definition the same electroweak radiative
corrections as the vector ones.  Small differences that can result
from $q^2 \neq 0$ are included in the theoretical uncertainty or
evolution uncertainty of the form factors.

The vacuum polarization correction to the muon bound state
wavefunction \cite{Goldman:1973qu}
must be individually evaluated
for different nuclei.  A detailed calculation gives
\begin{equation}
\RC({\rm H})_2 = 1.73{\alpha\over \pi} \simeq 0.004,
\label{eq:20}
\end{equation}
which is somewhat smaller than found by Goldman \cite{Goldman:1973qu}.
In the case of helium, we obtain
\begin{equation}
\RC({\rm He})_2 = 2.92{\alpha\over \pi} \simeq 0.0068.
\label{eq:21}
\end{equation}
Overall, we find
\begin{eqnarray}
1+\RC({\rm H}) &=& 1.028(4), \nn
1+\RC({\rm He}) &=& 1.030(4), 
\label{eq:22}
\end{eqnarray}
which modify the capture rate predictions in 
Eqs.~(\ref{eq:11}) and
(\ref{eq:26}-\ref{eq:27}) to 
\begin{eqnarray}
\Gamma(\captH)^{\rm singlet}_{1S} &=& 711.5(3.5)_{g_A}(3)_{RC} 
                                  (1-0.0108\delta g_P)^2 /\sec,\nn
\Gamma\left( \captHe \right)^{\rm EPM}_{\rm stat} &=& 1537 (22) /\sec,
\nn
\Gamma\left( \captHe \right)^{\rm IA}_{\rm stat} &=& 1506(8)(7)_{g_A}
(6)_{\rm RC}
 (1-0.013\delta g_P)/\sec.
\label{eq:23}
\end{eqnarray}
For hydrogen, comparison with the experimental results in
Eq.~(\ref{eq:3})  leads to $\delta g_p =-0.9\pm 1.2$,
\begin{eqnarray}
g_P^{\rm exp} &=& 7.3\pm 1.2 \qquad {\rm hydrogen}. 
\label{eq:24}
\end{eqnarray}
The electroweak radiative corrections have increased the value of
$g_P^{\rm exp}$ by about $+21\%$.  They bring theory and experiment
into agreement.  That situation is to be contrasted with the world
average $g_P^{\rm exp}=10.5\pm 1.8$ obtained \cite{Gorringe:2002xx}
from muon capture on hydrogen before the new MuCap result
\cite{Andreev:2007wg} and our evaluation of the radiative
corrections. On its own, that previous world average would have been
shifted to $g_P^{\rm exp}=11.7 \pm 1.8$ by the radiative corrections,
about a $2 \sigma$ deviation from the chiral perturbation theory
prediction.  However, including the MuCap \cite{Andreev:2007wg}
result, one finds the new world average from muon capture on hydrogen,
$g_P^{\rm exp}=8.7\pm 1.0$, in good agreement with the theoretical
prediction of Eq~(\ref{eq:10}).

For helium, radiative corrections spoil somewhat the good agreement
between experiment and the EPM prediction.  The new disagreement
suggests a smaller value of $g_A(-0.954m_\mu^2)_{{\rm He}\to {\rm H}}$
is likely or a significantly larger $g_P$ by about 25\% in magnitude
beyond PCAC predictions (a situation similar to hydrogen if we had
used the pre MuCap capture rates).  On the other hand the IA approach
\cite{Marcucci:2001qs} fares much better, leading to $g_P^{\rm exp} =
8.7\pm 0.6$ which is also in good agreement with chiral perturbation
theory.

In summary, when our calculation of the electroweak radiative
corrections to muon capture on hydrogen is combined with a new singlet
$\mu^-p$ capture rate measurement, it leads to $g_P^{\rm exp} =7.3\pm
1.2$ which is in very good accord with the prediction of chiral
perturbation theory, $g_P^{\rm theory} =8.2 \pm 0.2$.  That
agreement would seem to close a confusing chapter in nuclear physics
which has seen decades of disagreement regarding the value of
$g_P^{\rm exp}$.  It will be very interesting to watch continuing
improvements in the MuCap results.

\begin{acknowledgments} 

We thank the MuCap Collaboration for discussions and advising us about
their results. 
A.C. was supported by the Science and Engineering Research Canada.
W.J.M. acknowledges support by DOE grant DE-AC02-76CH00016. 
The work of A.S. was supported in part by 
the National Science Foundation Grant PHY-0245068. 
A.C. and W.J.M. thank the Institute for Nuclear Theory at the
University of Washington for its hospitality and the Department of
Energy for partial support during the completion of this work. 
\end{acknowledgments}

%\bibliographystyle{../../Tables/Archive/prsty}
%\bibliography{../../Tables/Archive/phd}

\end{document}